# Over 6 *µ*m thick MOCVD-grown Low-background Carrier Density ($10^{15}$ cm$^{-3}$) High-Mobility (010) *β*-Ga$_2$O$_3$ Drift Layers

Arkka Bhattacharyya[1,a], Carl Peterson[1,*], Kittamet Chanchaiworawit[1,*], Saurav Roy[1], Yizheng Liu[1], Steve Rebollo[1], and Sriram Krishnamoorthy[1,a]

[1]MaterialsDepartment, University of California Santa Barbara, Santa Barbara, California, 93106, USA.

*(\* these authors contributed equally)*

[a)] Email: arkkabhattacharyya@ucsb.edu, sriramkrishnamoorthy@ucsb.edu

*Abstract*: This work reports high carrier mobilities and growth rates, simultaneously in low unintentionally-doped UID ($10^{15}$ cm$^{-3}$) MOCVD-grown thick *β*-Ga$_2$O$_3$ epitaxial drift layers, with thicknesses reaching up to 6.3 *µ*m, using triethylgallium (TEGa) as a precursor. Record high room temperature Hall mobilities of 187-190 cm$^2$/Vs were measured for background carrier density values of 2.4-3.5×$10^{15}$ cm$^{-3}$ grown at a rate of 2.2 *µ*m/hr. A controlled background carrier density scaling from 3.3×$10^{16}$ cm$^{-3}$ to 2.4×$10^{15}$ cm$^{-3}$ is demonstrated, without the use of intentional dopant gases such as silane, by controlling the growth rate and O$_2$/TEGa ratio. Films show smooth surface morphologies of 0.8-3.8 nm RMS roughness for film thicknesses of 1.24 - 6.3*µ*m. Vertical Ni Schottky barrier diodes (SBDs) fabricated on UID MOCVD material were compared with those fabricated on hydride vapor phase epitaxy (HVPE) material, revealing superior material and device characteristics. MOCVD SBDs on a 6.3 *µ*m thick epitaxial layer show a uniform charge vs depth profile of ~ 2.4×$10^{15}$ cm$^{-3}$, estimated $\mu_{drift}$ of 132 cm$^2$/Vs, breakdown voltage (V$_{BR}$) close to 1.2 kV and a surface parallel plane field of 2.05MV/cm without any electric field management – setting record-high parameters for any MOCVD-grown *β*-Ga$_2$O$_3$ vertical diode to date.

## Manuscript

Advances in solid state power switching devices become increasingly important as the world transitions into a new era of electrification and the power electronics industry ventures beyond the Si power device roadmap. Wide-bandgap materials, such as SiC and GaN, are leading this revolution with demonstrated device efficiency and system-level performance advantages beyond Silicon's theoretical limits [1]. *β*-Ga$_2$O$_3$ is an emerging ultra-wide bandgap (UWBG) semiconductor that has gathered worldwide attention because of its intrinsic material properties such as a large bandgap (4.5–4.9 eV) and a large breakdown field strength (projected value of 8 MV/cm;) that are attractive for high-voltage switching devices. β-Ga$_2$O$_3$ is the only UWBG semiconductor that can have high quality melt-grown bulk substrates with dislocation densities as low as $10^4$ cm$^{-2}$ [2–8]. This enables substrate size scalability offering an affordable cost-competitive platform for the emerging Ga$_2$O$_3$ device demonstrations and faci its widespread progress. With the availability of shallow dopants, the n-type conductivity of Ga$_2$O$_3$ is easily tunable over many orders of magnitude leading to rapid experimental success in high voltage devices [9–18]. β-Ga$_2$O$_3$-based devices with V$_{BR}$ over 10 kV and breakdown electric fields exceeding the theoretical limits of SiC and GaN (experimental values: 5–7 MV/cm) have been demonstrated - establishing β-Ga$_2$O$_3$ as the most promising candidate material for next-generation very high voltage power-switching applications [19–40].

With a very high breakdown field and achievable low-background carrier densities, β-Ga$_2$O$_3$-based devices are well poised to deliver devices with tens of kilovolts blocking voltage capabilities [41]. For most of the high-voltage high-power applications that require voltages >3kV combined with currents >100A, lateral devices are not preferred due to the need for sizable chip areas [42]. Vertical device structures, however, present a highly desirable alternative, allowing for superior current drives, terminal current scalability, and more efficient field termination. The successful scaling of blocking voltages in Ga$_2$O$_3$ vertical devices relies on the detailed understanding and control of doping in the devices' drift layers. Rapid progress has been achieved in thick Ga$_2$O$_3$ drift layer growths by many different epitaxial methods such as HVPE, MOCVD, and LPCVD [43–57]. Amongst these, HVPE remains the dominant choice for high voltage vertical device demonstrations so far because it can be used to grow lightly doped thick (> 10 µm) drift layers. The commercial availability of > 2-inch HVPE-grown epitaxial layers has resulted in rapid progress in vertical device demonstrations. Films grown using MOCVD with triethylgallium (TEGa) as the Ga precursor showcase state-of-the-art crystal quality, material purity, and carrier mobility, typically achieving growth rates of ≤1.0μm/hr [58–65]. Trimethylgallium (TMGa), with its higher vapor pressure and faster reaction kinetics compared to TEGa, enables even faster growth rates of 10-15 μm/hr [49,51]. While MOCVD growth of $β$-$Ga_2O_3$ films using TMGa is relatively new and less explored, it faces challenges related to background C incorporation which negatively impacts the material purity and electronic transport. MOCVD-grown films using TEGa are more established but lack experimental studies for fast growth rates and vertical device demonstrations, with most studies limited to lateral FETs. Because of the β-hydride elimination process of the organic ethyl group during the growth reactions, TEGa-grown material tends to have lower unintentional C incorporation resulting in high material purity, doping control, and exceptional transport properties [52,66]. The primary focus of this study is to explore the highest device grade $β$-$Ga_2O_3$ films with high mobility, low doping, and the highest achievable thicknesses and growth rates using TEGa, beyond what has been demonstrated in the literature.

The $β$-$Ga_2O_3$ films were grown using a cold wall MOCVD Agnitron Agilis 100 reactor from Agnitron Technology Inc., USA that has a vertical quartz tube with a far injection showerhead. Films were grown on diced 5×5 $mm^2$ Fe-doped semi-insulating and Sn-doped conducting (010) native $β$-$Ga_2O_3$ substrates, which were commercially acquired from Novel Crystal Technology, Inc., Japan. All the films reported here were unintentionally n-doped $β$-$Ga_2O_3$ homoepitaxial films involving varying growth temperatures from 850 to 880 °C and chamber pressures from 5 to 60 Torr. Ga was introduced via TEGa precursor, and ultra-high pure $O_2$ gas (5N) was used as the oxygen source. High purity Argon (5N) was used as the carrier gas with flow rates of 2000 sccm. The flow rate of $O_2$ was adjusted from 400 to 1200 sccm, and the molar flow rate of TEGa ranged from 15 to 180 μmol/min to vary the growth rates and also the VI/III ratio. Before loading the substrates in the MOCVD reactor, they underwent a cleaning process involving acetone, isopropyl alcohol (IPA), and deionized water followed by a 49% HF dip of ~ 30 mins [65].

The growths were performed in the mass-transport limited regime at a growth temperature range of 850-880°C and the growth rates (GR) were scaled by increasing the TEGa molar flows. As shown in Figure 1(a), the GR monotonically increases with increasing TEGa molar flow while keeping the rest of the growth parameters nominally the same. Scaling the Ga molar flow from 15 to 180 μmol/min resulted in a proportional increase in GR from 0.3 to 4.5 $μ$m/hr. The film thickness was verified using cross-sectional SEM imaging and also using in-situ reflectance spectroscopy on c-plane sapphire substrates. The thickness on sapphire substrates aligned well with measurements of homoepitaxial $Ga_2O_3$ films, confirmed through high-voltage CV measurements, as discussed later. Figure 1(b) shows the effect of chamber pressure on the GR with a fixed Ga molar flow of 121 μmol/min and temperature of 880°C. It was observed that with increasing chamber pressure the GR decreased and the films ended up with lots of pits and particulates. This is expected because the increased chamber pressure increases the amount of gas phase pre-reactions [48]. Although higher chamber pressure enhances adatom species' diffusion length on the surface by reducing desorption and enables smoother films, it needs to be balanced to avoid excessive particulate formation. It is also important to have the best combination of chamber pressure and VI/III ratio for a particular GR to maintain a smooth surface morphology. Figure 1(c) shows the change in GR with varying $O_2$ flow rates from 400 to 800 sccm with a fixed Ga molar flow of ~ 121 μmol/min and growth temperature of 880°C. In the current growth window, an excess amount of $O_2$ has been provided (the $O_2$/TEGa molar ratio ≥ $10^2$), which ensures an $O_2$-rich growth condition. Consequently, the change in oxygen amount had a negligible effect on the GR, as depicted in Figure 1(c).

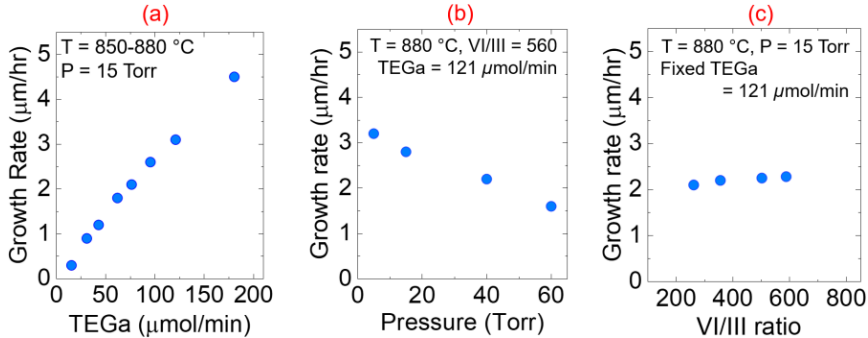

FIG. 1. Growth rate (GR) as a function of different growth parameters. (a) GR scaling from 0.3 to 4.5 $μ$m/hr with the scaling of TEGa flow rates. (b) GR variation with chamber pressure for a fixed TEGa flow rate of 121 $μ$mol/min, (c) GR shows little to no variation with different VI/III ratios (fixed TEGa and varying $O_2$) indicating that GR is still dictated by TEGa flow rate when temperature and pressure are nominally the same.

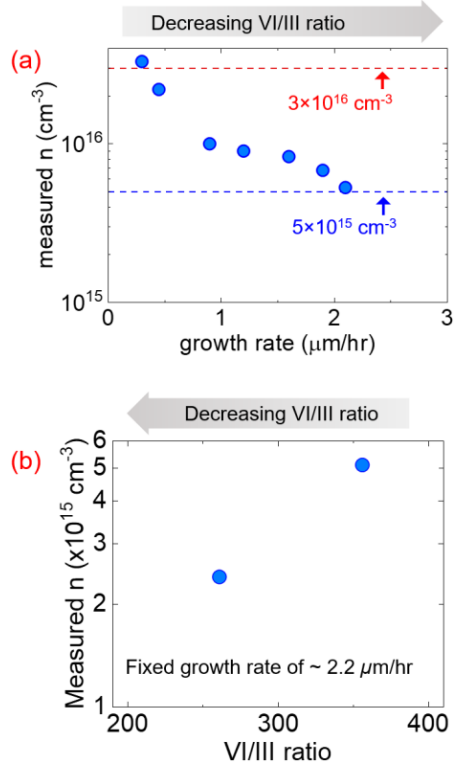

FIG. 2. Hall measured charge as a function of both GR and VI/III ratio. (a) Shows the decrease in Hall measured electron density ($n$) as the GR is increased and the VI/III ratio is decreased simultaneously. (b) Shows almost 2× reduction in $n$ when the VI/III ratio is halved keeping the GR fixed.

It is observed that while the $O_2$/TEGa (VI/III) ratio does not significantly impact the growth rate with a fixed TEGa flow rate, it does influence the background carrier densities in these unintentionally doped $Ga_2O_3$ films. Alternatively, fixing $O_2$ flow, temperature, and pressure while scaling the Ga molar flow inherently results in an increase in GR and a decrease in the VI/III ratio simultaneously. In Figure 2(a), the GRs are scaled from 0.3 to 2.2 $\mu$m/hr, while the VI/III ratio decreases from 2250 to 360. Beyond a GR of 3 $\mu$m/hr and VI/III precursor ratio of less than 200, the films become excessively rough on the surface or exhibit high electrical resistance, hindering electrical measurements. Background electron densities were measured using Hall measurements and later validated by CV measurements. The background electron densities for films with different GRs and VI/III ratios are shown in Figure 2(a). It can be seen that the background electron density reduces from $3.3\times10^{16}$ cm$^{-3}$ to $5.2\times10^{15}$ cm$^{-3}$ with a reduction in the VI/III ratio. While further detailed growth experiments are necessary to fully understand this, it is speculated that this effect may be attributed to background Si whose concentration and/or incorporation efficiency is altered with increasing GR, decreasing VI/III ratio, and higher carrier gas flows. The optimal combination for background carrier density, electron mobility, and surface roughness was found for a GR of 2.2 $\mu$m/hr and growth temperature of 880°C. Further reductions in VI/III for the same GR of 2.2 $\mu$m/hr result in a decrease in background electron density from $5.2\times10^{15}$ cm$^{-3}$ to $2.4\times10^{15}$ cm$^{-3}$ when the VI/III ratio is reduced from 360 to 260, as depicted in Figure 2(b). Films with thicknesses ranging from 1.24 $\mu$m to 6.3 $\mu$m exhibit RMS roughness values of 0.8 to 3.8 nm, demonstrating fairly smooth surfaces despite the high GRs and thicknesses as shown in Figure 3.

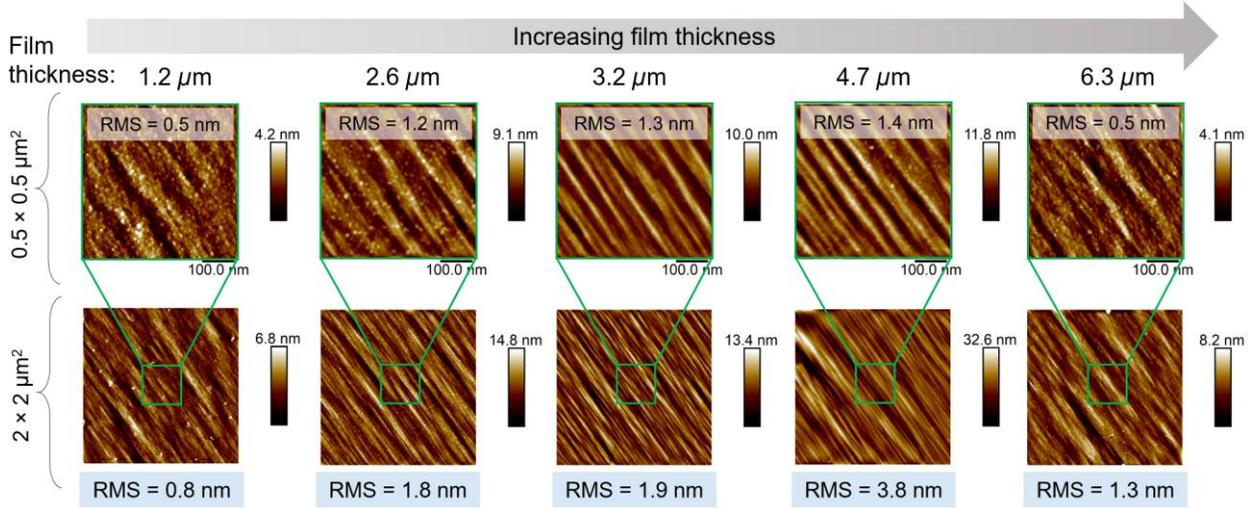

FIG. 3. 0.5×0.5 $\mu m^2$ (top row) and the corresponding 2×2 $\mu m^2$ area scans (bottom row) of the high growth rate films with different thicknesses.

Isolated van der Pauw structures were fabricated on thick epitaxial films grown on insulating (010) Fe-doped $\beta$-$Ga_2O_3$ substrates for Hall effect measurements. For isolation, $BCl_3$-based dry etching was used to achieve a total etch depth of 3 to 7 $\mu m$ for samples with film thicknesses ranging from 1.2 to 6.3 $\mu m$ (mesa depths extending through the films into the substrate). The epitaxial layers were completely removed in the etched regions to ensure optimal isolation, as confirmed by negligible leakage current below the measurement tool noise floor (0.1 nA/mm) between two isolation patterns. Further details can be found in the supplementary information. Figure 4 shows the room temperature Hall mobility for films with background electron density in the range of $2.4\times10^{15}$ to $9.5\times10^{15}$ $cm^{-3}$ with mobility values ranging from 190 to 168 $cm^2/Vs$, respectively. These mobility values are among the highest reported in the literature for this carrier density range. Literature reports with carrier concentration below $10^{16}$ $cm^{-3}$ involve intentional or unintentional compensation resulting in lowered mobility [67]. The high mobility values measured in this work suggests negligible compensation, which needs to be confirmed with the correlation of SIMS, Hall measurements, CV characterization and temperature-dependent Hall Effect characterization.

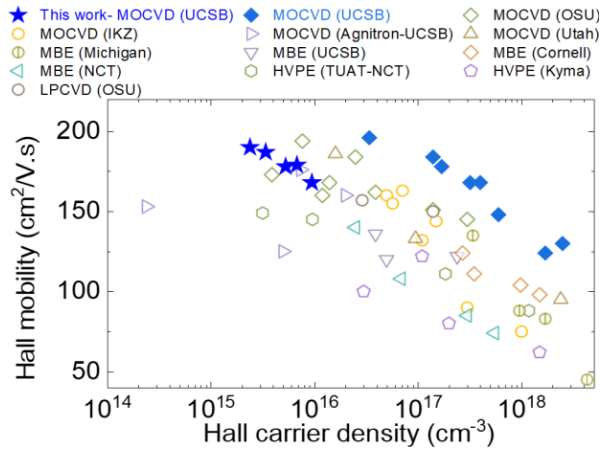

FIG. 4. Measured room temperature Hall mobility of the high GR MOCVD-grown films as a function of Hall carrier density, benchmarked with various state-of-the-art reports (MOCVD: UCSB [65], OSU [58,59], IKZ [46–48,63,68], Agnitron [50,60,61,67], Utah [64]) (MBE: Michigan [69], UCSB [15,70], Cornell [71], NCT [11], HVPE: TUAT-NCT [43], Kyma [45], LPCVD OSU [54]).

To further probe the charge vs. depth profile, we performed high voltage CV measurements to achieve punchthrough in Pt/Al$_2$O$_3$/Ga$_2$O$_3$ vertical MOS structures for drift layers on Sn-doped conducting substrates using a Keysight B1505 parameter analyzer with a capacitance measurement unit. MOS structures were used as they exhibit reduced reverse leakage in comparison to SBD structures during CV measurements. First, a ~ 70 nm Al$_2$O$_3$ is deposited on the thick film using thermal ALD process at 300°C. A metal stack of Pt/Au/Ni (30/100/50 nm) was evaporated and lifted off to form large ~ 1×1 mm$^2$ size top anode contacts. A metal stack of Ti/Au (50/200 nm) was evaporated on the back of the conducting substrate to form the backside Ohmic contact of the vertical structure. A large anode size is selected so that MOS structure has a minimum capacitance value of 10 pF or higher when the films are completely depleted at high reverse bias (after achieving punchthrough). This combination of dielectric and a large anode size facilitated more accurate CV measurements by eliminating leakage at high reverse bias and maintaining capacitance values well above the noise floor of the measurement tool. We characterized two of our thickest films labeled as MOCVD A (4.7 $\mu$m thick) and MOCVD B (6.3 $\mu$m thick). To help with the perspective, we have also included a commercially acquired HVPE grown low-doped epilayer on (001) conducting Sn-doped substrate. We also fabricated bare planar Ni SBDs on all three samples to observe the effect of the epilayer on device parameters. All the parameters extracted from vertical MOS and SBDs are summarized in Figure 5, 6 and Table I.

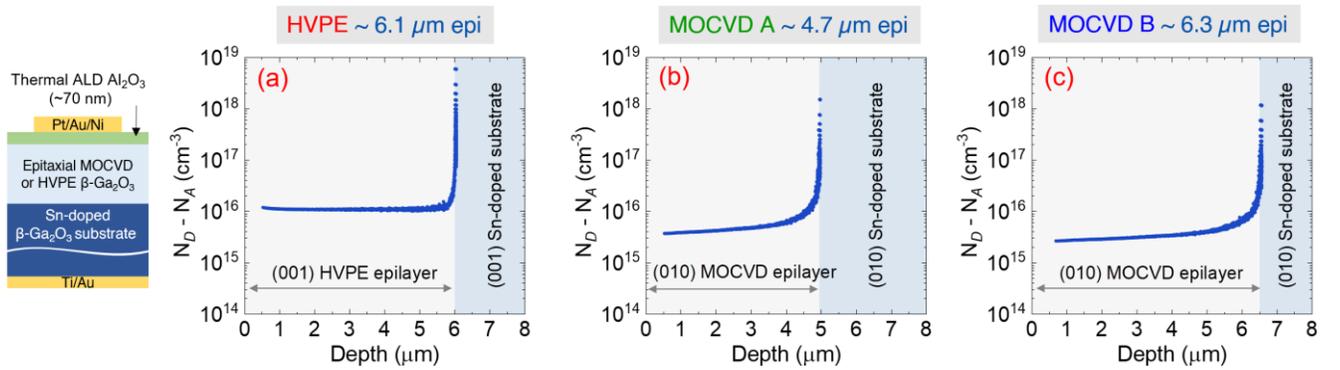

FIG. 5. Extracted apparent charge vs depth profile from high-voltage CV measurements for the (a) HVPE, (b) MOCVD A and (c) MOCVD B samples. The schematic shows the structure of the vertical MOS structure used for these measurements.

The charge density profile extracted from capacitance-voltage (CV) measurements is shown in Figure 5 (AC signal frequency of 0.2 to 1 MHz). The CV characteristics showed little to no change within this frequency range. The voltage was swept from zero bias to high reverse bias until full epilayer depletion or punchthrough was achieved. The experimentally extracted oxide capacitance was later measured from accumulation capacitance and subtracted from the total capacitance to correct for depth extraction calculations. The CV-extracted thicknesses of the epilayers matched closely with those obtained from X-section SEM images on co-loaded sapphire pieces. This agreement validates the thickness measurements through two independent methods. The charge distribution showed a nearly flat profile. Sample MOCVD A which is a 4.7$\mu$m thick epilayer shows an extracted $N_D - N_A$ value of 3.5×10$^{15}$ cm$^{-3}$. The MOCVD B sample has a thickness of 6.3$\mu$m and the extracted $N_D - N_A$ was 2.4×10$^{15}$ cm$^{-3}$. These numbers match very well with carrier densities extracted from Hall measurements, again confirming the low carrier concentration in the MOCVD-grown β-Ga$_2$O$_3$ layers. Low background impurity is desired for vertical power devices to reach high breakdown voltages and breakdown fields simultaneously.

Table I: This table summarizes and compares the different film properties extracted from Hall measurements, high voltage CV measurements, and electrical measurements on Ni SBDs on MOCVD-grown and commercially acquired HVPE-grown $\beta$-$Ga_2O_3$ films.

| Properties | Film thickness ($\mu$m) [a] | $N_D-N_A$ ($cm^{-3}$) [b] | RT Hall Mobility ($cm^2$/Vs) [c] | Ni SBD ||||||
|---|---|---|---|---|---|---|---|---|---|
| | | | | SBH (eV) | ideality factor ($\eta$) | $R_{on,sp}$ (m$\Omega\cdot cm^2$) | Estimated $\mu_{drift}$ ($cm^2$/Vs) | $V_{BR}$ (V) | $E_{\|surface}$ @ $V_{BR}$ (MV/cm) |
| HVPE | 6.1 | $1.1\times10^{16}$ | - | 1.03 | 1.08 | 3.9 | 104 | 594 | 1.58 |
| MOCVD A | 4.7 | $3.5\times10^{15}$ | 187 | 1.12 | 1.18 | 6.6 | 123 | 686 | 1.63 |
| MOCVD B | 6.3 | $2.4\times10^{15}$ | 190 | 1.28 | 1.24 | 9.1 | 132 | 1186 | 2.05 |

[a] Measured from high voltage CV measurements. MOCVD films were also measured by X-section SEM on coloaded sapphire substrates.
[b] Measured from high voltage CV measurements on drift layers grown on conducting substrates
[c] Hall measurements were done on drift layers grown on co-loaded Fe-doped semi-insulating substrates

Current–voltage (J–V) characteristics for the planar SBDs were compared. The devices were circular Schottky pads with a diameter of 100$\mu$m. A metal stack of Ni/Au/Ni (30/100/50 nm) was evaporated and lifted off to form the Schottky metal anodes. Figure 6(a) shows and compares the ON-state and OFF-state performance of the SBDs on the three epilayer samples. The Schottky barrier heights (SBH) were extracted from the CV measurements. The ideality factors ($\eta$) were extracted using the thermionic emission model. The differential specific on-resistance ($R_{on,sp}$) for the diodes was extracted by normalizing the forward current ($I_{on}$) with the effective anode diameter considering a 45° current spreading angle. The drift mobility can be estimated from the $R_{onsp}$, as $\mu_{drift} = \left(\frac{L_{drift}}{qn(R_{on.sp}-R_{sub})}\right)$, where q is the fundamental electron charge, n is the free carrier density, $L_{drift}$ is the epilayer thickness, and $R_{sub}$ is the substrate resistance which is estimated to be 0.6-0.8 m$\Omega\cdot cm^2$ from the substrate datasheet. The extracted SBH, $\eta$, $R_{onsp}$, and $\mu_{drift}$ are summarized in Table I. All the diodes show near ideal ON-state characteristics with high rectification ratio of $> 10^9$ for an applied voltage range of $\pm$ 2V. The HVPE diode has a slightly lower SBH of 1.03 eV compared to MOCVD diodes (SBH = 1.1-1.3 eV) which is expected as (010) orientation is observed to have a comparatively higher SBH in literature when compared to the (001) orientation [72]. The $R_{onsp}$ with current spreading is extracted to be 3.9, 6.6 and 9.1 m$\Omega\cdot cm^2$ for the HVPE (epi thickness = 6.1$\mu$m), MOCVD A (epi thickness = 4.7$\mu$m) and MOCVD B (epi thickness = 6.3$\mu$m), respectively. The extracted $\mu_{drift}$ values are also comparatively better in the MOCVD material ($\mu_{drift}$ = 124-132 $cm^2$/Vs) compared to the HVPE material ($\mu_{drift}$ = 104 $cm^2$/Vs). The higher $\mu_{drift}$ could be due to lower free carrier densities in MOCVD films compared to the HVPE drift layer.

Reverse J–V characteristics and breakdown characteristics of the planar non-field plated Ni SBDS on the three samples are shown in Figure 6 (b) and compared in Table I. The breakdown in each of the diodes was found to be catastrophic in nature. So, in this work, we are considering $V_{BR}$ as the catastrophic breakdown voltage. The HVPE diode gave a breakdown voltage of ~ 594V. The MOCVD A and MOCVD B diodes gave breakdown voltages of 686V (epi thickness = 4.7$\mu$m) and 1186 V (epi thickness = 6.3 $\mu$m), respectively. It is to be noted that a $V_{BR}$ close to ~ 1.2kV is the highest $V_{BR}$ ever demonstrated on an MOCVD-grown vertical $Ga_2O_3$ diode [73–75]. All the diodes attained punchthrough well before the catastrophic breakdown. The punchthrough voltages were measured from high-voltage CV measurements. The measured reverse bias punchthrough voltages for the three diodes were 402, 141 and 169V for the HVPE, MOCVD A and MOCVD B samples, respectively. This matches very well for the expected punchthrough voltages analytically which are calculated to be 379V, 110 and 143 V, respectively. This corroborates well with the thickness and free carrier density of the reported films. Using the punchthrough field profile, the surface parallel plane fields ($E_{\|surface}$) at the center of the anode were calculated to be 1.58, 1.68 and 2.05MV/cm for the HVPE, MOCVD A, and MOCVD B, respectively. It can be seen that the MOCVD diodes have higher surface fields compared to the HVPE sample. It is noteworthy that a surface field of 2.05 MV/cm can be achieved in the MOCVD SBD without any field management - the highest ever measured in an MOCVD-grown vertical diode. It can be said the low-doped (010) MOCVD material performs well in terms of $V_{BR}$ and reverse leakage characteristics and is comparable to the commercial HVPE material. This indicates the superior electronic grade of the MOCVD material in terms of forward conduction (low $R_{on,sp}$ and high estimated $\mu_{drift}$) as well as high E-field handling capabilities at high reverse biases. Further growth studies are required to understand the evolution of device-killing defects that could form at higher growth rates and cause leakage-related premature breakdown as seen in the literature, which is important for current scaling in vertical power devices.

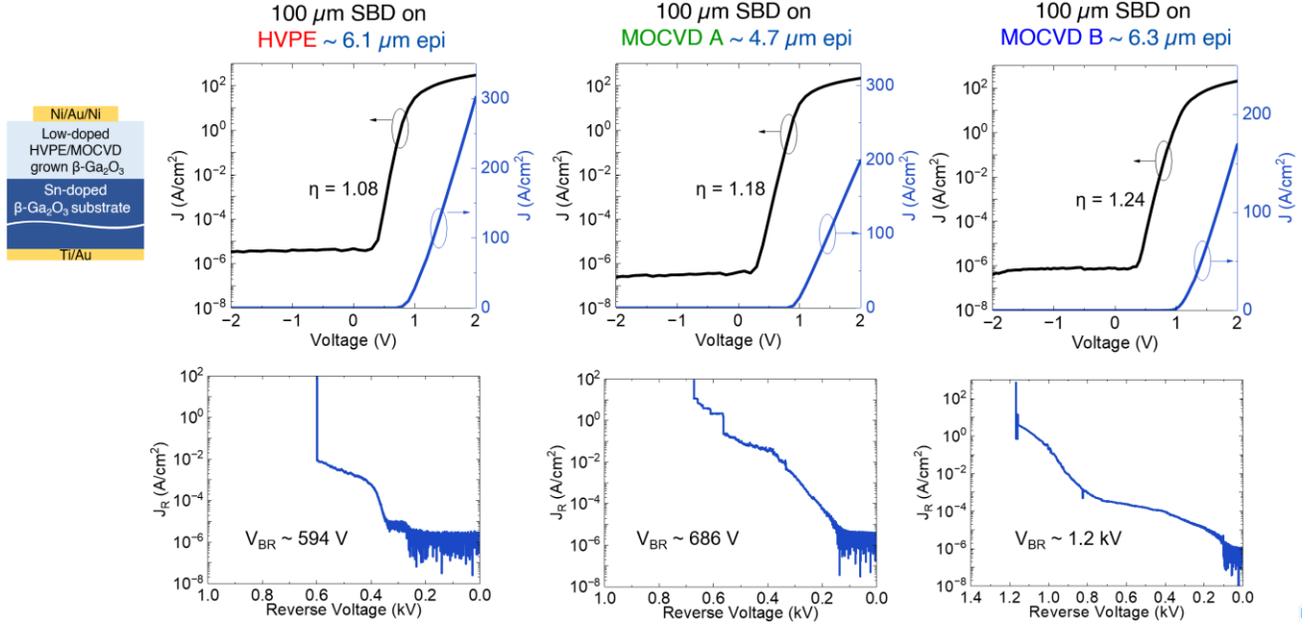

FIG. 6. J-V characteristics of the SBDs fabricated on the HVPE, MOCVD A and MOCVD B samples (top row) and their corresponding reverse breakdown characteristics (bottom row). The schematic shows the structure of the vertical SBD structure used for these measurements.

In summary, we demonstrate high-quality electronic device grade $\beta$-Ga$_2$O$_3$ films with high Hall and drift mobilities, low $10^{15}$ cm$^{-3}$ background carrier densities, and high epilayer thicknesses of up to 6.3 μm, all at a notably high growth rate of 2.2 μm/hr. This was accomplished using TEGa in a cold wall far-injection showerhead vertical MOCVD reactor – both thickness and growth rate are significantly higher than what has been demonstrated in literature. Although very high growth rates up to 4.5 μm/hr were achieved using TEGa, the highest electronic device grade material is grown using a growth rate of 2.2 μm/hr. A controlled background carrier density scaling from $3.3 \times 10^{16}$ cm$^{-3}$ to $2.4 \times 10^{15}$ cm$^{-3}$ is demonstrated by carefully tuning the growth rates and VI/III ratio during growth. Record high room temperature Hall mobilities of 187-190 cm$^2$/Vs are achieved for background carrier density values of $2.4$-$3.5 \times 10^{15}$ cm$^{-3}$, grown at a growth rate of 2.2 μm/hr. The films exhibit remarkably smooth surface morphologies with RMS roughness ranging from 0.8 to 3.8 nm for film thicknesses of 1.24-6.3 μm, facilitating immediate device fabrication post-growth. Comparison between vertical Ni SBDs fabricated on MOCVD material and commercially available HVPE material reveals superior material and device characteristics. A high voltage CV technique is demonstrated, enabling non-destructive electrically measured charge profile probing throughout the drift layer for microns of thicknesses leading to an accurate estimation of parallel plane breakdown fields. MOCVD SBDs on a 6.3 μm thick epitaxial layer show a fairly uniform charge vs. depth profile of $\sim 2.4 \times 10^{15}$ cm$^{-3}$ with an estimated drift mobility of 132 cm$^2$/Vs, a $V_{BR}$ close to 1.2 kV, and a surface parallel plane field of 2.05 MV/cm without any electric field management – setting record-high parameters for any MOCVD-grown $\beta$-Ga$_2$O$_3$ vertical diode to date. The demonstrated results mark a significant advancement in low-doped $\beta$-Ga$_2$O$_3$ thick drift layer epitaxy, a crucial element in the development of high-voltage power devices, and emphasizes the pivotal role of MOCVD in the future progress of vertical $\beta$-Ga$_2$O$_3$ power devices.

# ACKNOWLEDGMENTS

This work was supported in part by the Air Force Office of Scientific Research, under Award No. FA9550-21-1-0078 (Program Manager: Dr. Ali Sayir), and in part by the Coherent/II-VI Foundation Block Gift Program. A portion of this work was performed in the UCSB Nanofabrication Facility, an open access laboratory.

# Over 6 μm thick MOCVD-grown Low-background Carrier Density ($10^{15}$ cm$^{-3}$) High-Mobility (010) *β*-Ga$_2$O$_3$ Drift Layers

Arkka Bhattacharyya[1,a], Carl Peterson[1,*], Kittamet Chanchaiworawit[1,*], Saurav Roy[1], Yizheng Liu[1], Steve Rebollo[1], and Sriram Krishnamoorthy[1,a]

## Supplementary Information

**Van der Pauw structure fabrication:**

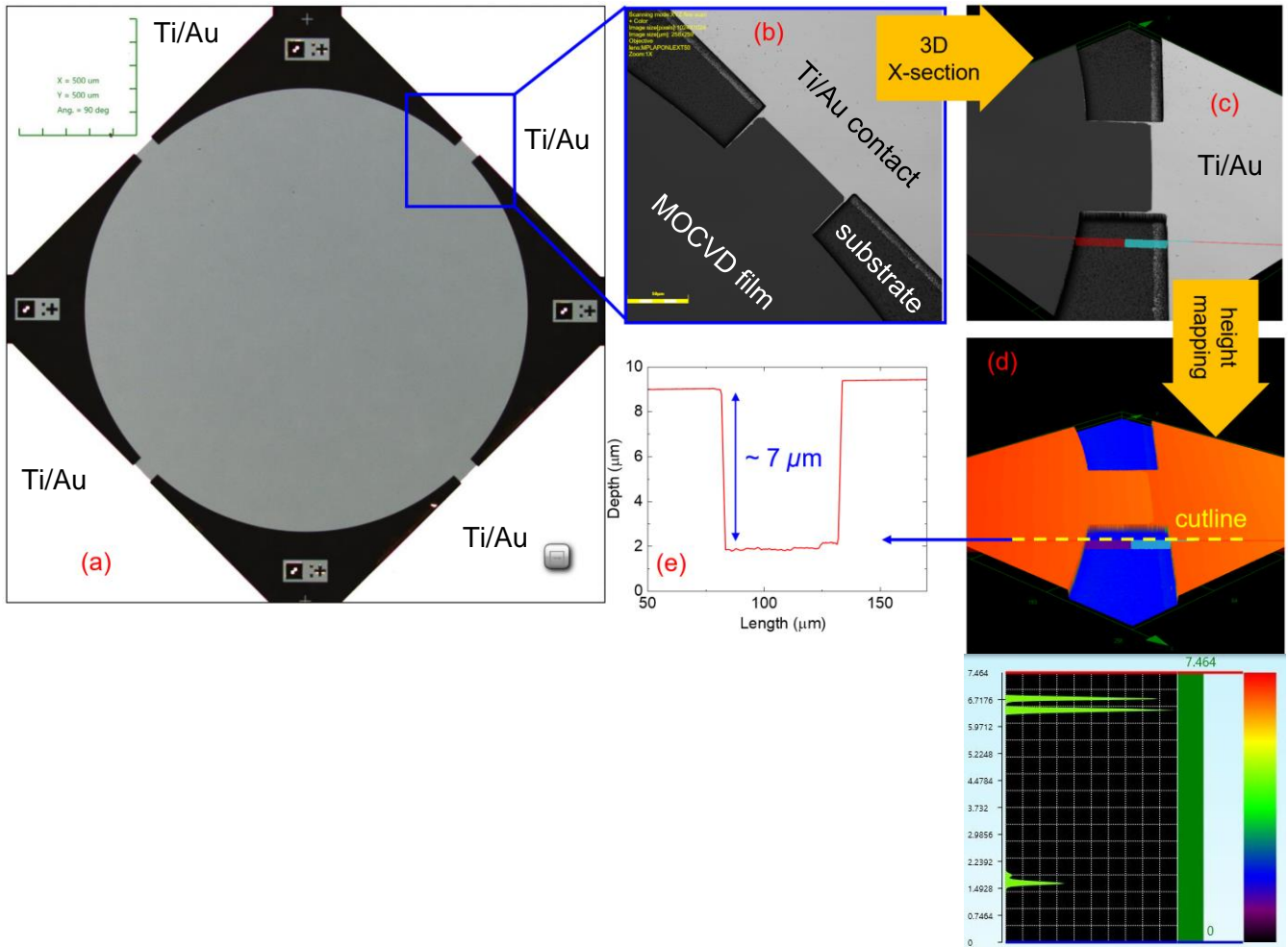

Figure S1: Confocal Laser Scanning Microscope images of the deep-mesa etched van der Pauw structures fabricated on a 6.3 μm MOCVD-grown homoepitaxial film on Fe-doped (010) semi-insulating Ga$_2$O$_3$ substrate. (a) Shows a top-view image of the van der Pauw structure. (b) Magnified image of the "point" contact for current injection utilizing a large periphery Ti/Au Ohmic contact. The contact size to distance ratio is 0.04. (c) Tilted perspective of the "point" contact to show the mesa etching. (d) height map of the mesa etched region. (d) cutline showing the measured etch depth of ~ 7μm which is deeper than the epilayer thickness.

Processing: A Ni/SiO$_2$ (500nm/600nm) hard mask was used to define the van der Pauw structure. The thick Ni metal provided enough selectivity for etching microns of Ga$_2$O$_3$ using a ICP BCl$_3$ dry etching with an estimated Ga$_2$O$_3$ etch rate of 90 nm/min.

## Large area vertical MOS CV structures:

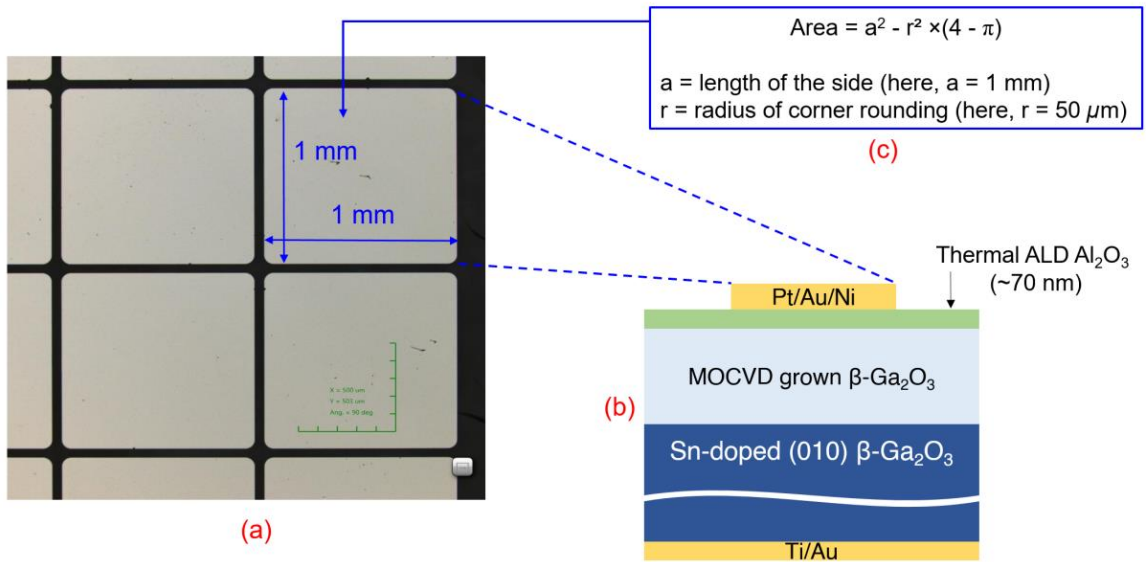

Figure S2: Shows the structure used to high voltage MOS CV measurements on MOCVD and HVPE-grown drift layers on conducting Sn-doped substrates. (a) shows the top view microscope images of the corner rounded large "squircle" shape CV pads. "Squircle" shape avoids sharp corners (E-field peaking) as well as gives better area utilization. (b) cross-section schematic of the MOS structure with $Al_2O_3$ dielectric. (c) the analytical formula used to calculate area of the 1 mm square with corner rounding radius of 50 μm.

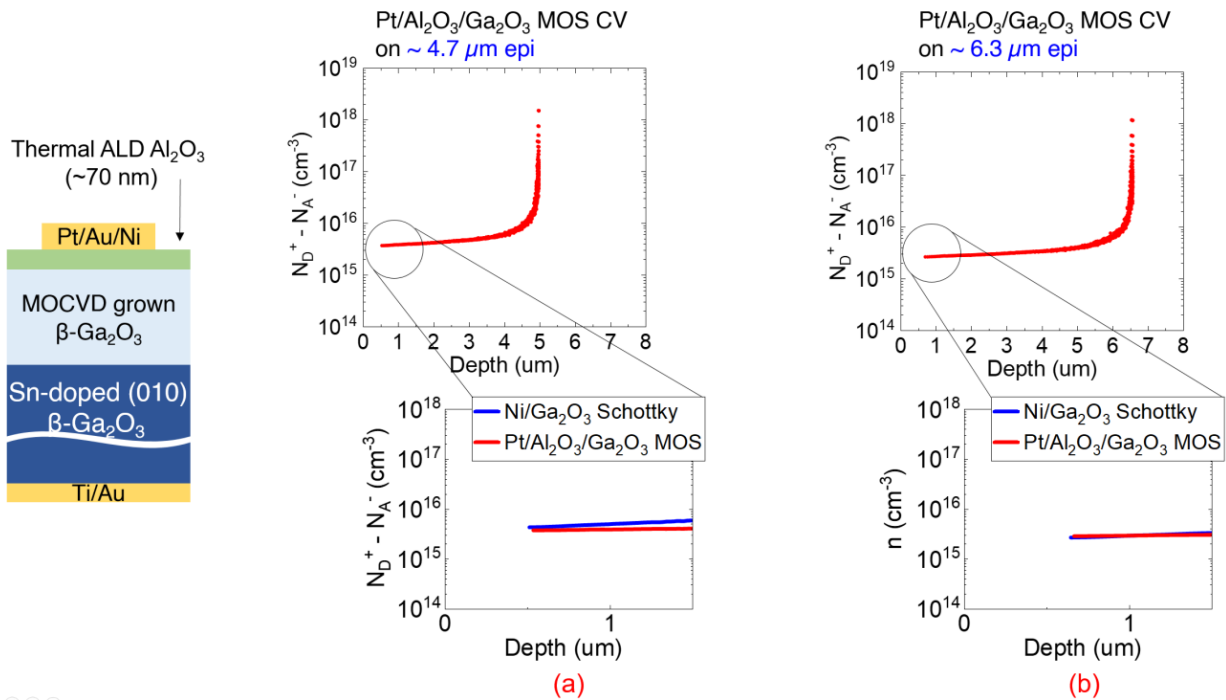

Figure S3: Comparison of the MOS and Schottky contact CV extracted charge vs depth. The small area Schottky resulted in very small measured capacitances of < 2 pF and high reverse leakage after the films were completely depleted at high reverse bias that resulted in erroneous depth profile extraction. Nevertheless, the Schottky contacts were able to accurately probe the charge profile for the top 1-2 μm of the drift layers. Figure shows the extracted charge profile of the (a) MOCVD A sample and (b) MOCVD B sample where the bottom images show that the MOS process did not alter the charge profile as it matches very well with the Schottky results.

**Large area optical microscope images:**

1.5 mm

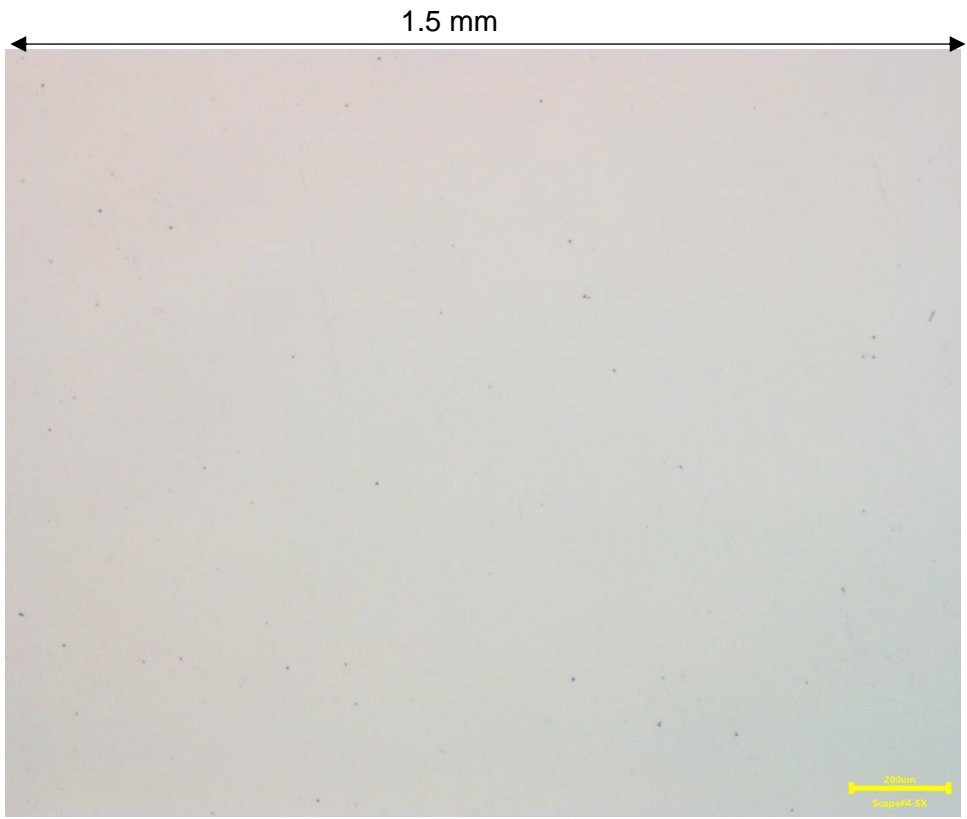

(a) Optical microscope top view image showing the surface of a MOCVD-grown film with a thickness of 1.24 $\mu$m.

1.5 mm

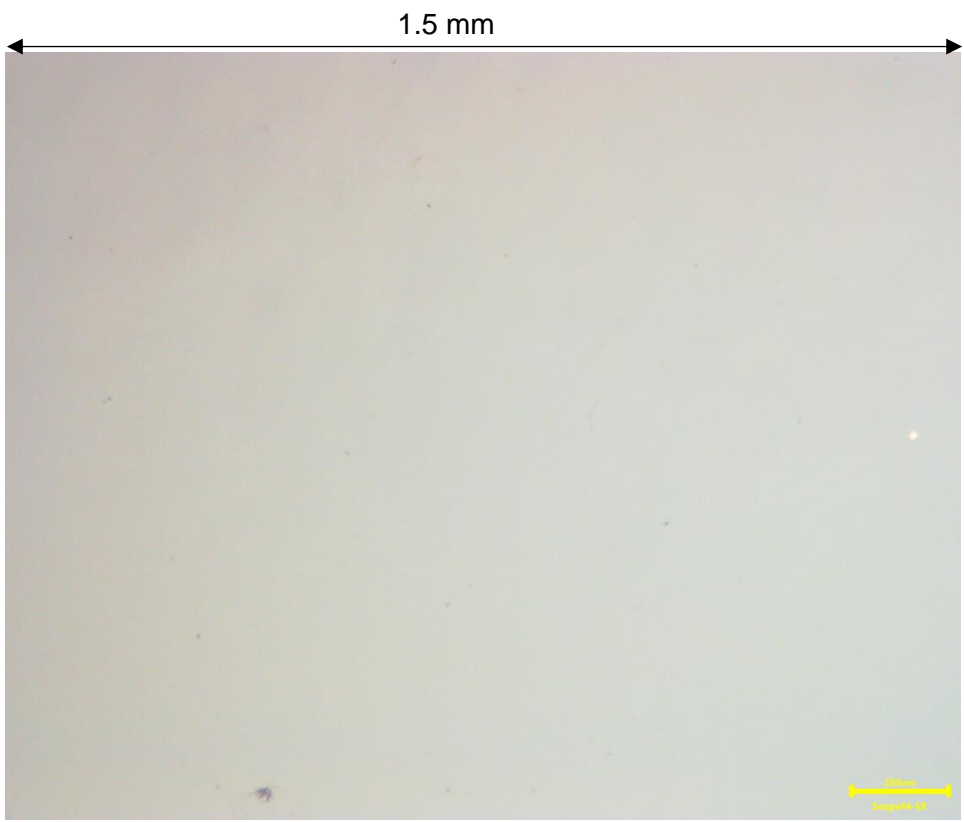

(b) Optical microscope top view image showing the surface of a MOCVD-grown film with a thickness of 2.1 $\mu$m.

1.5 mm

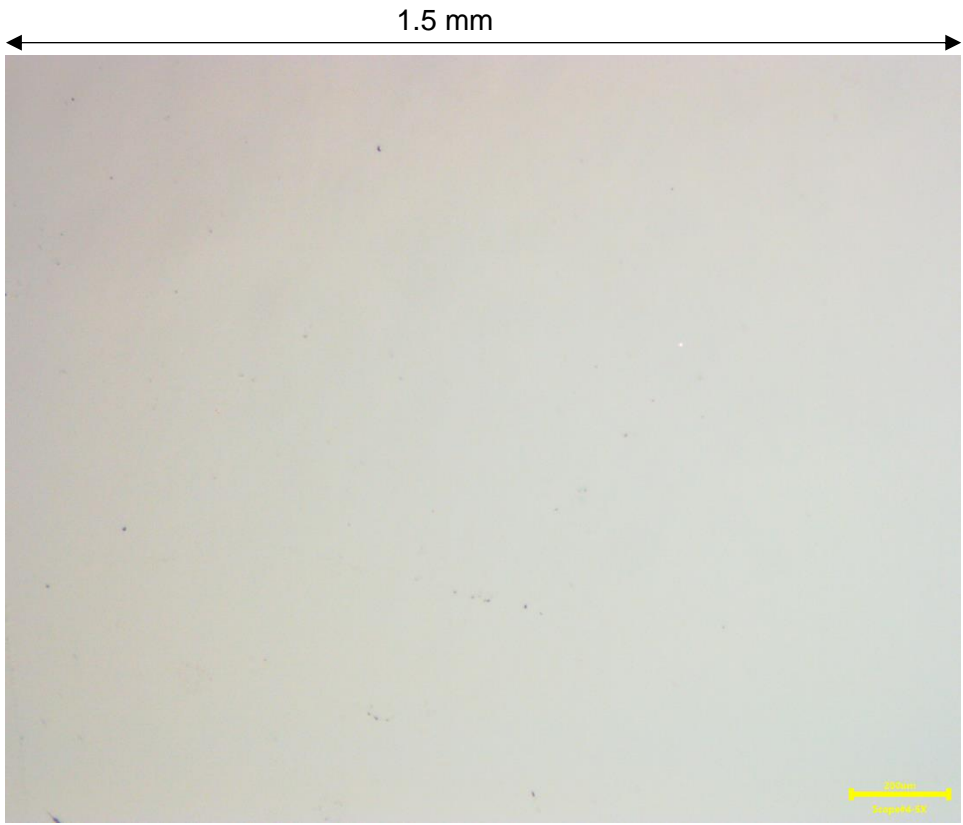

(c) Optical microscope top view image showing the surface of a MOCVD-grown film with a thickness of 2.6 μm.

1.5 mm

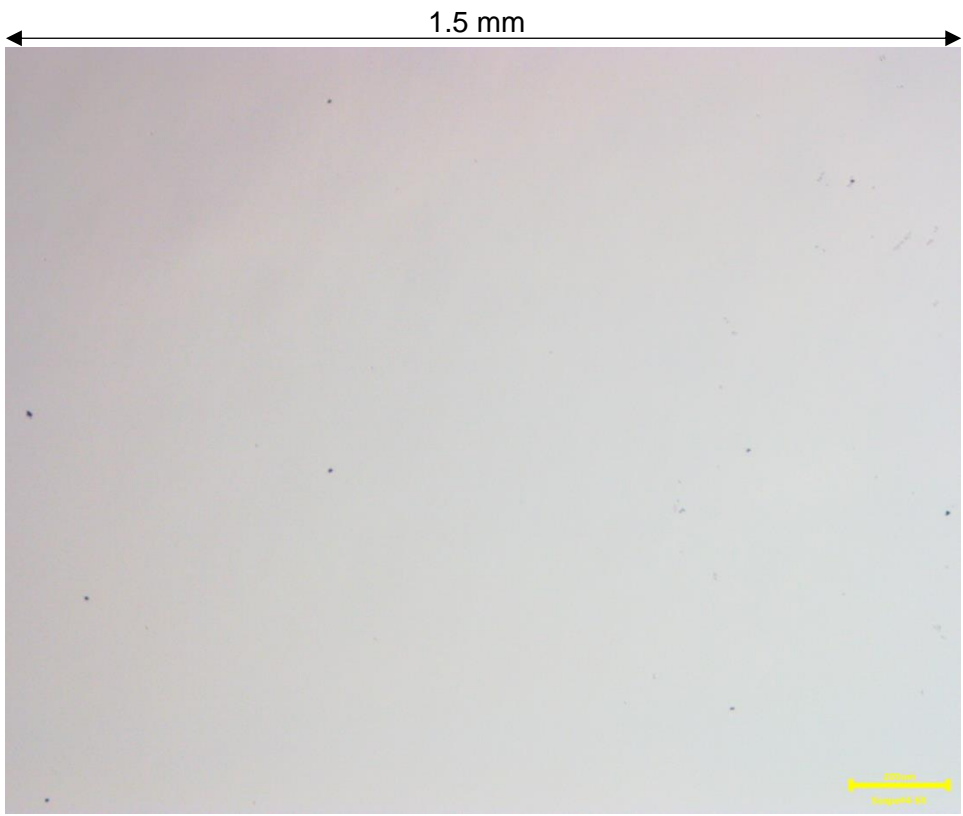

(d) Optical microscope top view image showing the surface of a MOCVD-grown film with a thickness of 4.7 μm.

1.5 mm

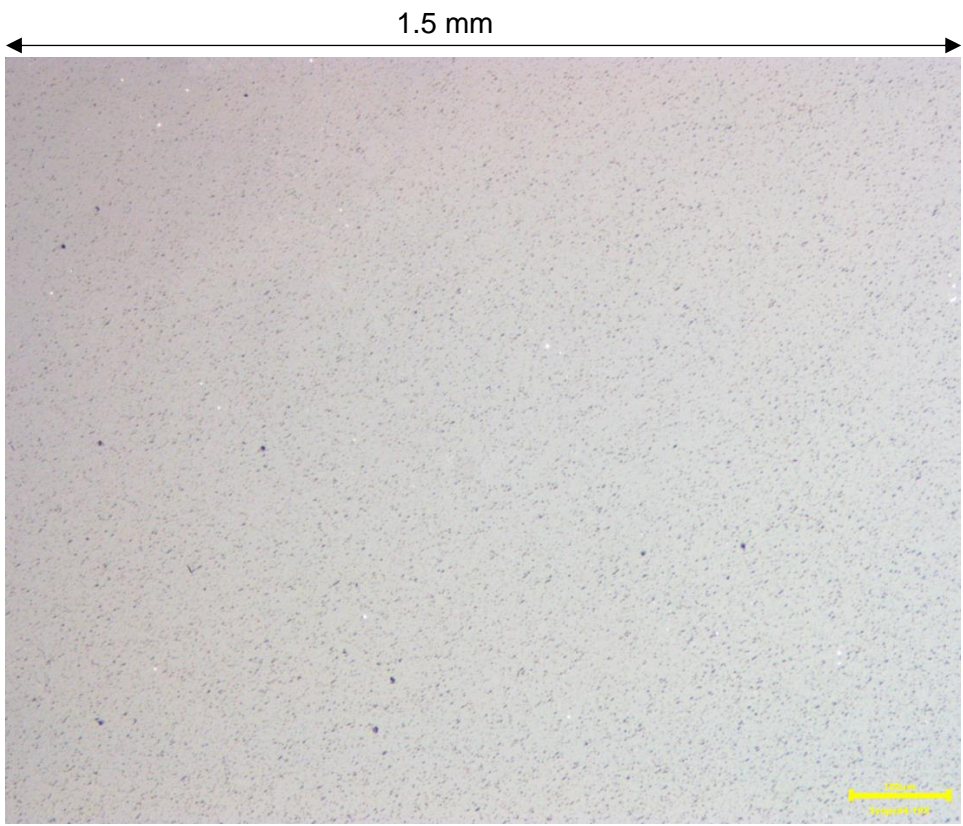

(e) Optical microscope top view image showing the surface of a MOCVD-grown film with a thickness of 6.3 μm.